# Causality principle in reconstruction of sparse NMR spectra

*Maxim Mayzel, Krzysztof Kazimierczuk, and Vladislav Yu. Orekhov* \*

***Abstract***: *Rapid development of sparse sampling methodology offers dramatic increase in power and efficiency of magnetic resonance techniques in medicine, chemistry, molecular structural biology, and other fields. We suggest to use available yet usually unexploited prior knowledge about the phase and the causality of the sparsely detected NMR signal as a general approach for a major improvement of the spectra quality. The work gives a theoretical framework of the method and demonstrates notable improvement of the protein spectra reconstructed with two commonly used state-of-the-art signal processing algorithms, compressed sensing[1] and SIFT[2].*

The invention of multidimensional magnetic resonance (MR) experiments 40 years ago led to success of the modern MRI and NMR spectroscopy in medicine, chemistry, molecular structural biology, and other fields. The approach, however, has an important weakness: the detailed site-specific information and ultimate resolution obtained in two and higher dimensional experiments are contingent on the lengthy data collection needed for systematic uniform sampling of the large multidimensional spectral space. Days or even years of data collection, required to achieve the optimal resolution, are often incompatible with stability of the studied system or economical considerations. A fundamental solution for this problem stems from an observation that upon appropriate transform, e.g. from the NMR time to frequency domain, MR signal becomes nearly-black or sparse, i.e. essentially zero in the vast majority of points and thus largely redundant. Darkness of the MR images and NMR spectra is a key for the remarkable success and rapid development of the non-uniform (or compressed) sampling (NUS) methods[1, 3, 4-6]. The darker is an object, the less experimental measurements are needed for its recovery[7]. In the field of MRI, large variety of sparsifying transforms were adopted from image processing or were newly invented for obtaining the darkest representation of the MRI signal.

In NMR, the Fourier transform connects complex free induction decay (FID) signal and frequency spectrum. A properly phased spectrum consists of the real absorption part used for analysis and the imaginary dispersion part. Since an absorption signal is much narrower than the dispersion, the latter contribute the most to the spectrum brightness. The main result of this paper is the notion that for many of the currently used algorithms, e.g. compressed sensing[1, 5], SIFT[2], maximum entropy[4, 8], MINT[6], etc., quality of the NMR spectrum obtained from the NUS data is limited by the dispersion signal. We show that the causality property of the NMR signal can be used to construct sparsifying transform, which eliminates the dispersion part and allows spectrum reconstruction with better fidelity and from fewer measurements.

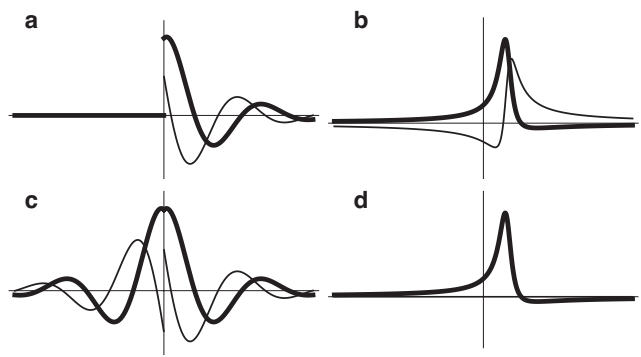

*Figure 1.* (**a**) FID and (**c**) virtual-echo representations of the NMR time domain signals with the corresponding spectra (**b** and **d**, respectively). Real and imaginary parts are shown in bold and thin lines, respectively. Note that the spectrum in panel (**d**) has zero imaginary part. Small zero order phase $0.15\pi$ is used to illustrate the effect of non-zero phase on the signal in the time and frequency domains.

It is well known that the Fourier transform of a causal time signal $s(t)$ leads to spectrum, whose real and imaginary parts can be produced from each other using the Kramers–Kronig relations also known as the Hilbert transform [9]. Specifically for NMR signal, the causality principle means that the FID signal is only observed after excitation of the spin system, e.g. by a radiofrequency pulse, and is zero before the excitation. The Kramers–Kronig relations are illustrated in Figure 1. Causal FID signal $S_{FID}(t)$ (Fig. 1a) and spectrum in Fig. 1b are related via the Fourier transform. Spectrum in Fig. 1d is produced from the one in Fig. 1b by zeroing its imaginary part. The inverse Fourier transform of real spectrum in panel d gives a complex time domain signal (Fig. 1c), whose real and imaginary parts are essentially even and odd parts of the real and imaginary components of the FID (Fig 1a), respectively. Thus, signal in Fig. 1c is produced by time reversal and complex conjugate of the FID.

$$S_{VE}(t) = \begin{cases} S_{FID}(t) & t \geq 0 \\ S_{FID}^{*}(-t) & t < 0 \end{cases} \quad (1)$$

[*] M. Mayzel, V. Yu. Orekhov
Swedish NMR Centre
University of Gothenburg
Box 465, S-405 30 Göteborg, Sweden
Fax: (+46)31 786 3886
E-mail: vladislav.orekhov@nmr.gu.se

K. Kazimierczuk
Centre of New Technologies
University of Warsaw
Żwirki i Wigury 93, 02-089 Warsaw, Poland

Acknowledgements: Swedish Research Council (grant 2011-5994); Swedish National Infrastructure for Computing (grant SNIC 001/12-271); Polish Ministry of Science and Higher Education (grant IP2011 023171), National Centre of Science (grant DEC-2012/07/E/ST4/01386); Foundation for Polish Science, TEAM programme; Dina Katabi and Haitham Hassanieh (Dept Electr Eng & Comput. Sci., MIT) for an inspiring discussion; an unknown reviewer for pointing to Kramers–Kronig relations.

The original signal $S_{FID}(t)$ can be obtained from $S_{VE}(t)$ by zeroing the signal for negative time. In the following, we call $S_{VE}(t)$ signal in Eq. 1 virtual-echo (VE). Direct transition from panel d to panel b in Fig. 1 is done by the Hilbert transform. In practice, the Hilbert transform algorithm takes the detour d → c → a → b in order to use the computationally efficient fast Fourier transform.

The spectrum (Fig. 1d) obtained from the VE representation (Fig. 1c) consists of the traditionally looking real part and zero imaginary part. Depending on the phase, the real part can contain absorption, dispersion, or mixture of the both modes (see Appendix). Given *a priori*, the signal phase, Eq. 1 allows us to obtain the pure absorption representation of the spectrum and, thus, to construct a sparsifying transform that produces significantly darker spectrum than the traditional Fourier transform of the original FID. As it is shown in the Appendix, Eq. 1 can be generalized for a spectrum of any dimensionality.

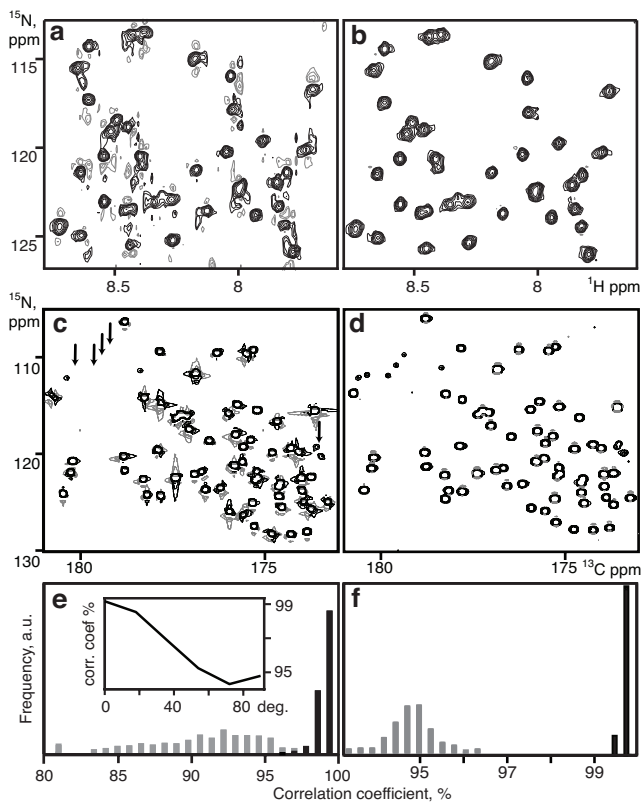

*Figure 2.* Comparison of SIFT (**a**,**b**,**e**) and CS (**c**,**d**,**f**) spectral reconstructions obtained using time-domain signal in traditional FID (**a**,**c**) and VE (**b**,**d**) presentations. (**a**,**b**) 2D $^1$H-$^{15}$N HSQC of ubiquitin (25% NUS). (**c**,**d**) $^{13}$C-$^{15}$N projection from a 3D HNCO spectrum of ubiquitin (0.7% NUS). In the pairs of spectra, the contours are shown at the same level. Arrows in panel (**c**) indicate several true weak signals present in the VE reconstruction (**d**) but missing in panel (**c**). Histograms (**e**,**f**) show distribution of the correlation coefficients between signal intensities measured in the reference spectrum and the spectra reconstructed with VE (black) and FID (grey) (**e**) SIFT: 500 resampling trials with 25% NUS and $^{15}$N signal support size of 60 Hz; (**f**) CS: 200 resampling trials with 0.7% NUS. Inset in panel (**e**) shows the median (over 50 resampling trials) of correlation coefficients for the VE processing versus uncorrected zero order phase.

Signal presentation related but not equivalent to Eq. 1 was suggested by Nagayama[10]. Without going into details irrelevant to our work, we only note that Nagayama's approach is not suitable for producing quantitative spectra (for details see[11]). While preparing this paper, we found that a signal representation similar VE was recently used in a new method for NUS spectra reconstruction[12]. However, there the VE was deeply entangled into the method-specific algorithm and general implications of the VE presentations for NUS spectra reconstruction were not discussed.

The virtual-echo time-domain presentation given by Eq. 1 should be distinguished from the traditional true-echo signal previously suggested for obtaining pure-absorption spectra[11, 13]. The former is a mathematically equivalent presentation of a regular FID, the latter requires a specialized experimental setup. Both echo type approaches are capable of eliminating the broad dispersion signals and thus provide basis for efficient sparsifying transforms. The conventional MRI signal in the *k*-space corresponds to the true-echo, which contributes to success of the compressed sensing techniques in that field. However, to our knowledge, the true-echo has not been used so far in relation to the NMR spectra reconstruction from NUS data.

Obtaining NMR spectrum from a time-domain signal is a typical example of the mathematical inverse problem[14]. When all data points in the signal are present, the solution of the problem is trivial and is given by the DFT. In case of NUS, most of the data in the time-domain signal are missing and the unconstrained inverse problem has infinite number of solutions (i.e. spectra). A unique and "correct" spectrum is obtained by introducing additional assumptions such as maximum entropy, maximal sparseness, etc. The VE presentation is equally applicable to traditional fully sampled and NUS signals. When the former is processed using DFT, FID and VE presentations lead to the equivalent spectra as illustrated in Fig. 1. For NUS, however, path a → c → d in Figure 1 represents significant advantage over the traditional processing, which is a → b → d.

Figure 2 demonstrates benefits of the VE signal for two modern spectra recovering algorithms used for NUS signal: *Spectroscopy by Integration of Frequency and Time Domain* (SIFT)[2] and *Compressed Sensing by Iterative Reweighted Least Squares* (CS-IRLS)[1]. Both methods can be applied without modifications to either traditional FID or VE signal. With SIFT making use of the *prior* knowledge on positions of dark regions in a spectrum and CS-IRLS searching for the darkest among all possible spectra consistent with the measured data, both methods are expected to benefit from the darker representation of the spectrum provided by VE.

For a given number of NUS measurements, quality of the SIFT reconstruction improves, when the larger fraction of the spectrum area can be assumed to be free from signals and contain only the baseline noise. In our calculations, the signal-free area is defined by a mask, which excludes rectangles of a defined size around all peaks in the spectrum. This corresponds, for example, to a set-up in relaxation and kinetics studies[15], where the peak positions are known and only their intensities or integrals need to be defined. Figures 2a and 2b show reconstructions of a 2D $^1$H-$^{15}$N HSQC spectrum obtained using only 25% of the data from the full experiment. By avoiding broad dispersion peaks, the VE signal presentation allows smaller rectangles, i.e. larger fraction of the spectrum covered by the mask (Appendix Fig. A2), which leads to the much better spectrum (Fig 2b) and more accurate peak intensities in comparison to the SIFT reconstruction from the original FID (Fig 2e and Appendix Fig. A3). Figure 2e (inset) illustrates that any prior information about the signal phase reduces contribution of the dispersion part and thus produces spectrum of better quality. For most of multidimensional experiments, zero order phases for the indirect spectral dimensions are known and thus can be corrected in the time domain to values close to zero prior to the spectrum reconstruction. In the worst case of the uncorrected 90° phase of all signals, quality of the spectrum approaches that of the spectrum produced from the traditional FID representation.

Similarly to SIFT, CS also assumes, that the major part of a spectrum is dark. However, no assumption is made about the exact location of the dark regions, which creates an apparently unsolvable combinatorial problem. Yet, it has been recently reformulated as a relatively simple task of spectral $l_p$-norm ($0<p\leq1$) minimization[16]. Here we apply an IRLS algorithm[1] to reconstruct a 3D HNCO spectrum sampled at the level of 0.7 %, without VE (Fig 2c) and with VE in both indirect dimensions (Fig 2d). It can be seen, that VE improves the reconstruction significantly by providing better line shapes, more accurate peak intensities (Fig 2f), and revealing low intensity signals. The effect can be explained by the basic CS theorem, binding the number of properly reconstructed spectral points, which is essentially a measure of spectrum darkness, with the sampling level[16]. With VE, fewer points contribute to each peak in the spectrum and thus relatively low sampling level is sufficient to fulfil the condition for the successful CS reconstruction. Additionally, the application of the VE narrows the distribution of spectral magnitude. This gives advantage to the non-convex $l_p$-norms with $p<1$ over the $l_1$-norm in the CS algorithms and allows further reduction of sampling level[17]. It should be emphasized, that the striking advantage of the VE demonstrated in Fig 2, is mostly due to the very low sampling level. Without the VE, high quality reconstructions by CS and SIFT are also possible, but require at least twice as many sampling points for the presented spectra (Appendix Figs. A2,A4).

As it was pointed by Donoho, at al. [7] there is an unambiguous relation between the darkness of NMR spectrum and quality of the spectral reconstruction by the Maximum Entropy or minimum $l_1$-norm minimisation. It is therefore likely that most of related methods including FM-reconstruction[18], MINT[6], IstHMS [14], etc. will also benefit from the VE signal.

We show that the causality property of the NMR signal can be exploited to dramatically enhance performance of the CS, SIFT and probably many other algorithms commonly used for the reconstruction of NUS spectra. The reason for this is that the use of well-known Kramers–Kronig relations allows constructing more efficient sparsifying transformation of the time domain signal than the traditional Fourier transform of the complex FID. Our findings open a way to significant reduction in measurement time and improvement of the quality of NUS spectra and thus to increase of power and appeal of multidimensional NMR spectroscopy in multitude of its existing and future applications.

## *Experimental Section*

The fully sampled $^1$H-$^{15}$N HSQC spectrum of 0.2 mM human ubiquitin was recorded at 25 °C on 800 MHz Bruker Avance III HD spectrometer equipped with a TXI cryoprobe. The spectrum 98 complex points were acquired for the $^{15}$N dimension corresponding to the acquisition time of 34 ms. The NUS time-domain data was produced by selecting measurements in accordance with NUS schedules produced by the program *nussampler* from the *MDDNMR* software package[19] For processing of the spectrum, we implemented SIFT algorithm in MATLAB as described in the original paper[2]. For the Figure 2e region of spectrum between 8.8 and 8.3 ppm in $^1$H dimension was used for the processing. To measure intensities in the spectra reconstructed from VE with the uncorrected zero order phase (inset in Fig. 1e), the spectrum was phased after the reconstruction in the frequency domain using Hilbert transform. The 3D NUS HNCO spectrum of 1 mM human ubiquitin was acquired at 25 °C on 600 MHz Varian UNITY Inova spectrometer with a cold probe using 6.1% NUS out of the Nyquist grid of 120 x 79 points (43 ms and 28 ms) for the $^{13}$C and $^{15}$N spectral dimensions, respectively. The spectrum was processed at the sampling levels 0.25 - 6.1 % using CS-IRLS module[17] of the *MDDNMR* software.

## *Appendix*

**True and virtual echoes**

We illustrate relation between the virtual and true echo for the specific case of a single-component NMR signal with frequency $\Omega$, transverse relaxation rate $\alpha$, and phase $\phi$

$$S_{FID}(t) = \theta(t)\exp[-i\Omega t - \alpha t + i\phi] \quad -\infty < t < \infty \quad \text{(A1a)}$$

$$S_{VE}(t) = \exp[i\Omega t - \alpha|t| + i\phi \, sign(t)] \quad -\infty < t < \infty \quad \text{(A1b)}$$

where $\theta(t)$ is the Heaviside step function. Equation A1b, which is derived using Eqs. 1 is an equivalent presentation obtained from the original FID signal (Eq. A1a) without adding or loosing information.

In general case of $\phi \neq 0$, $S_{VE}(t)$ has a discontinuity at the time point zero. The true-echo signal, which is observed in the spin-echo type experiments[13, 20] and used in pseudo-echo signal processing[21]

$$S_{TE}(t) = \exp[-i\Omega t - \alpha|t| + i\phi] \quad -\infty < t < \infty \quad (A1c)$$

is continuous for any phase $\phi$. Fourier transform of the signals from equations A1 gives the following spectra:

$$F_{FID}(\omega) = \cos(\phi) L(\omega) + \sin(\phi) D(\omega) + i[\sin(\phi) L(\omega) - \cos(\phi) D(\omega)] \quad (A2a)$$

$$F_{VE}(\omega) = 2\cos(\phi) L(\omega) + 2\sin(\phi) D(\omega) \quad (A2b)$$

$$F_{TE}(\omega) = 2\cos(\phi) L(\omega) + 2i\sin(\phi) L(\omega) \quad (A2c)$$

where $L(\omega)$ and $D(\omega)$ are absorption and dispersion parts:

$$L(\omega) = \frac{1}{\sqrt{2\pi}} \frac{\alpha}{\alpha^2 + (\Omega - \omega)^2} \quad (A3a)$$

$$D(\omega) = \frac{1}{\sqrt{2\pi}} \frac{\Omega - \omega}{\alpha^2 + (\Omega - \omega)^2} \quad (A3b)$$

Signals and corresponding spectra given by Eqs. A1,A2,A3 are illustrated in Figures 1, A1. Equations A2 and A3 explain differences between spectra produced from the normal FID and echo signals. The former always contain equal amount of the absorption and dispersion, whose contributions to real and imaginary parts of the spectrum depend on the signal phase. Virtual-echo spectrum has zero imaginary part. Its real part contains mixture of absorption and dispersion parts. If the phase is known, $F_{VE}(\omega)$ can be adjusted to the pure absorption mode. The true-echo spectrum is purely absorptive, with the signal power distributed between the real and imaginary parts. Both the true and virtual echo approaches can be used to produce dark, pure absorption spectrum.

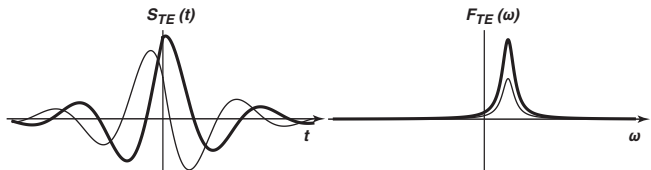

*Figure A1.* The time signals from Eq. A1c and the corresponding spectrum (Eq. A2c). Small $\phi = 0.15\pi$ is used to illustrate the effect of non-zero phase on the signal in the time and frequency domains.

**Digital considerations**

In practical NMR experiment, $N$ signal points are sampled at regular time intervals $\Delta$ over a period of time $t_{max} = \Delta(N-1)$. Considering that the discreet Fourier transform (DFT) assumes that the signal is periodic with the period $t_{max}$ and starts at the time point zero, equation for the virtual echo presentation of the digitized signal is:

$$S_{VE}(k\Delta) = \begin{cases} S_k, & 0 \leq k < N \\ S^*_{2N-k+\mu-1}, & N + \mu \leq k < 2N \end{cases} \quad (A4a)$$

Here and in the following, we consider two cases for position of the first measured data point: ($\mu = 1$) at time point $t = 0$, i.e. when all frequency components of the FID have the same phase, and ($\mu = 0$) at the half dwell time, $t = \Delta/2$. Eq. A4a can be seen as a sum of two signals: right-zero-padded (up to $2N$ points) FID and the complex conjugate of the left-zero-padded time reversed FID. For $\mu = 1$ and $\mu = 0$, VE signal is

$$S_{VE} = [(S_0 + S_0^*)/2, S_1, \ldots, S_{N-1}, 0, S_{N-1}^*, S_{N-2}^*, \ldots, S_1^*] \quad (A4b)$$

$$S_{VE} = [S_0, S_1, \ldots, S_{N-1}, S_{N-1}^*, S_{N-2}^*, \ldots, S_1^*, S_0^*] \quad (A4c)$$

respectively. If zero order phase $\phi$ of the signal is known, it should be applied to the time-domain FID signal prior to the conversion to the VE and spectrum reconstruction. Then, DFT of the signal in Eq. A4b gives a spectrum with pure absorption real part and zero imaginary part. With corrected zero order phase, spectrum of the signal in Eq. A4c contains only absorption contribution, which is however distributed between the real and imaginary parts. As in the case of the true-echo, this does not affect darkness of the spectrum and, if needed, the pure absorption spectrum with zero imaginary part can be obtained by the linear phase correction of $180^0$ in the frequency domain.

**Generalization for N-dimensions**

The virtual-echo transformation defined by Eq. A4 can be easily generalized for a spectrum of any dimensionality. For example, since the NUS is typically used only for the indirectly detected dimensions, for a three-dimensional spectrum we will consider a two-dimensional signal. For every time point in the 2D time domain ($0 \leq t_1 < t1_{max}$, $0 \leq t_2 < t2_{max}$), States method of NMR signal detection gives four real measurements, which we denote $R1R2$, $R1I2$, $I1R2$, and $I1I2$. From these we can build four complex matrices:

$$S_{\pm\pm}(t_1, t_2) = (R1 \pm iI1)(R2 \pm iI2) \quad (A5)$$

For example, complex matrix $S_{++}(t_1, t_2)$ is obtained as

$$(R1 + iI1)(R2 + iI2) = R1R2 - I1I2 + i[I1R2 + I2R1]$$

Then, we double sizes of the matrices in Eq. A5 by zero padding and perform the time-reverse operations for the individual dimensions similar to Eq. A4. Finally, after summation of the four resulting matrices we obtain a complex matrix of the 2D virtual-echo signal ($0 \leq t_1 < 2 t1_{max}$, $0 \leq t_2 < 2 t2_{max}$):

$$S_{VE}(t_1, t_2) =$$

$$\begin{bmatrix} S_{++}(t_1, t_2) & S_{+-}(t_1, t2_{max} - t_2) \\ S_{-+}(t1_{max} - t_1, t_2) & S_{--}(t1_{max} - t_1, t2_{max} - t_2) \end{bmatrix} \quad (A6)$$

The two-dimensional DFT of $S_{VE}(t_1, t_2)$ produces complex 2D spectrum with zero imaginary part. When phase of the signal is known *a priori* and corrected prior to the DFT, the spectrum is obtained in the pure absorption mode. As the final remark, we note that while the original FID for a multidimensional spectrum is hyper-complex, the VE signal presentation is always complex and, thus, directly amenable for processing using N-dimensional DFT, CS, and other algorithms dealing with complex signals.

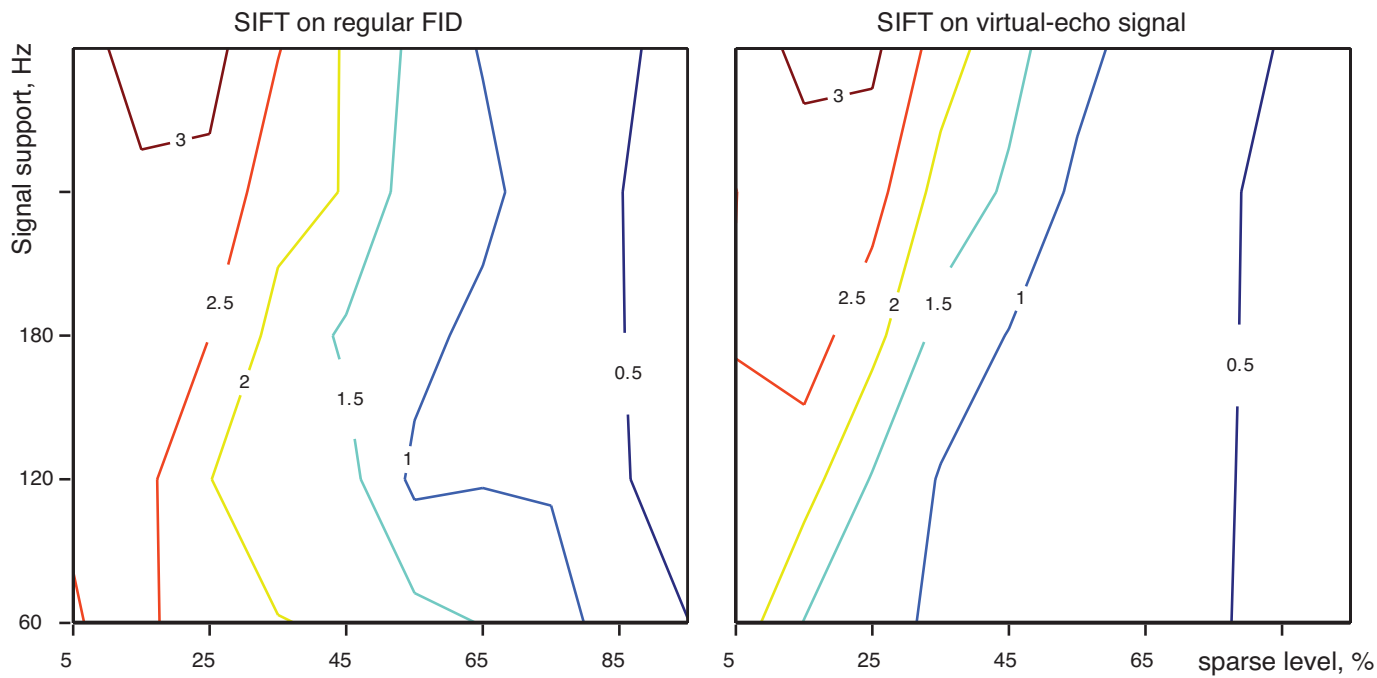

*Figure A2.* SIFT signal reconstruction quality scores for $^1$H-$^{15}$N HSQC spectrum of ubiquitin versus sampling level and size of the signal support mask for the $^{15}$N spectral dimension. The score is calculated as an RMSD of the difference between the reference spectrum (fully sampled, processed with DFT) and the SIFT reconstruction applied to regular FID (***left***) and to VE signal (***right***). The RMSD is calculated over all signal regions (manually verified peak maximum ± 37.6 Hz $^1$H and ±30 Hz $^{15}$N) and normalized to the noise level in the full reference spectrum. For every spectrum reconstruction, 10 resampling trials with different NUS schedules were performed and the reported value corresponds to the mean score. For any given sampling level, the VE allows tighter sample support mask and shows consistently lower (i.e. better) RMSD scores and, thus better reproduces the peak intensities and line shapes.

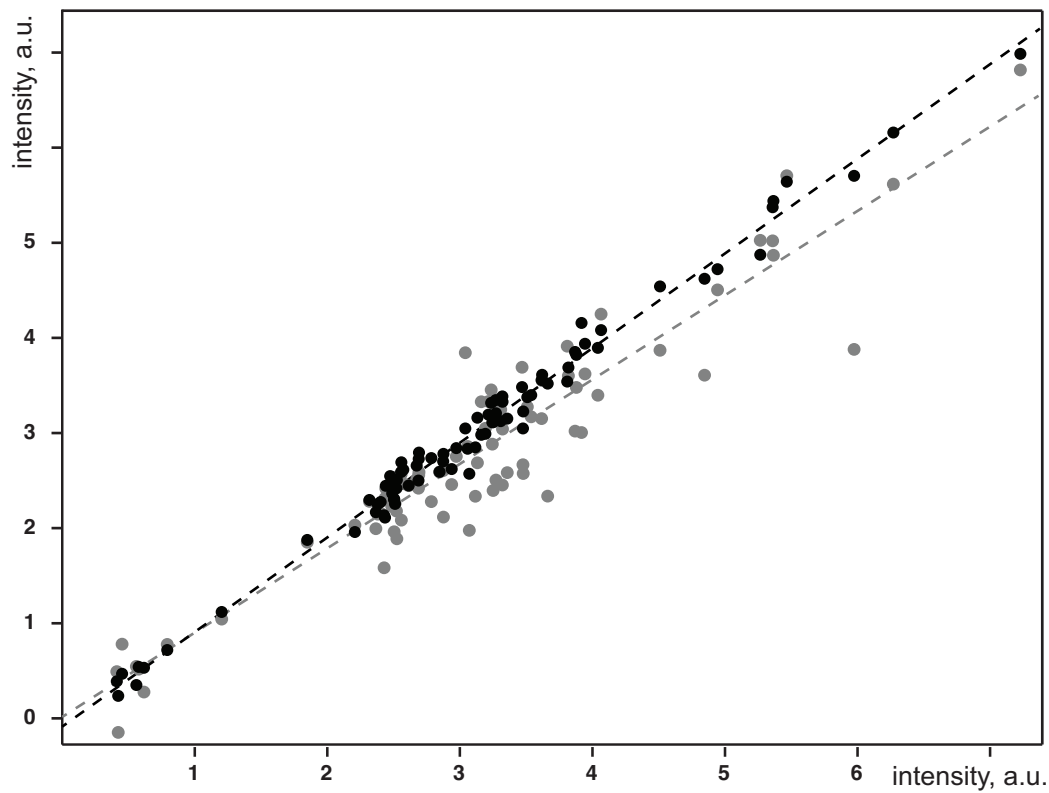

*Figure A3.* Accuracy of the peak intensities in the 2D $^1$H-$^{15}$N HSQC spectra of ubiquitin shown in Figures 2a and 2b in the main text. 25% NUS spectrum was processed with SIFT applied to the VE signal (***black***) or to regular FID (***grey***); in both cases size of the signal support region in $^{15}$N dimension was set to 60 Hz. Intensities of the peaks in the reconstructed spectra are shown against the corresponding intensities in the fully sampled reference spectrum. Lines correspond to linear regression fit. Correlation coefficients for the peak intensities are 0.99 and 0.94 for the SIFT reconstruction with and without VE, respectively.

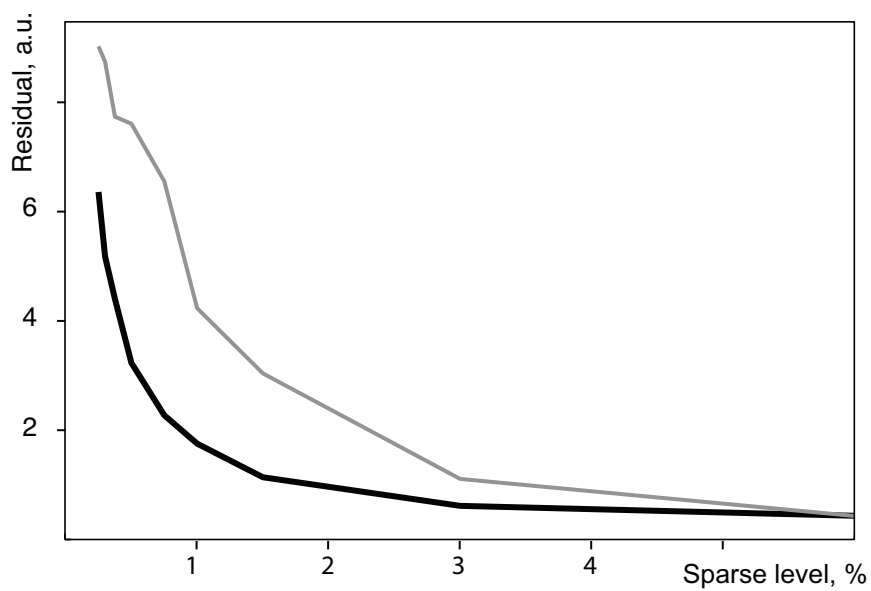

***Figure A4.*** Quality scores versus the sampling level are shown for 3D HNCO spectrum of ubiquitin reconstructed by CS-IRLS using regular FID (***grey line***) and VE signal representations (***black line***). The score is defined as an RMSD of the difference between the reference spectrum and the corresponding CS-IRLS reconstruction measured over the signal regions (±50 Hz in all spectral dimensions around every peak in a complete manually verified peak list). As the reference we use 6% NUS HNCO averaged over the reconstructions obtained with and without VE.